# Gyrokinetic simulations of the influence of electron cyclotron current drive on tearing mode instabilities in tokamaks

Jingchun Li[1], Chijie Xiao[1], Zhihong Lin[2], Dongjian Liu[3] and Xiaoquan Ji[4], Xiaogang Wang[5]

[1]FSC and State Key Lab of Nuclear Physics and Technology, Department of Physics, Peking University, Beijing 100871, People's Republic of China
[2]University of California, Irvine, CA 92697, USA
[3]Sichuan University, Chengdu 610064, People's Republic of China
[4]Southwestern Institute of Physics, Chengdu 610041, People's Republic of China
[5]Harbin Institute of Technology, Harbin 150001, People's Republic of China
Corresponding email: zhihongl@uci.edu

**Abstract**
A gyrokinetic simulation of the influence of electron cyclotron current drive and ion kinetic effect on the m/n=2/1 tearing mode (TM) instabilities is presented in HL-2A and DIII-D tokamak configurations. The TM evolution is calculated with a finite mass electron model and the rf current source is obtained by ray-tracing and the Fokker-Planck method. The TMs are found to be perfectly stabilized by a continuous 1MW 68GHz X2-mode in HL-2A tokamak, while instabilities in the DIII-D discharge (with lower value of CR$\equiv I_{rf}/I_0$, where $I_{rf}$ is the wave driven current and $I_0$ is the equilibrium plasma current) are only partially stabilized with the 1MW 110GHz X2-mode due to inadequate power input. The result also indicates that a helicon current drive is more efficient than a continuous ECCD. Analysis of the GTC simulation reveals, both in HL-2A and DIII-D, that the presence of ions can reduce the island width as well as the growth rate. Furthermore, the kinetic effect of thermal ions on TM is found to be more pronounced with higher ion temperature.

## 1. Introduction

Tearing mode instabilities degrade the plasma performance and even lead to the plasma disruptions.[1-3] So far, various methods for TM control have been established, such as electron cyclotron current drive (ECCD) [4], lower hybrid current drive (LHCD) [5], externally applied resonant magnetic perturbations [6-7], and neutral beam injection (NBI) [8-9]. Since ECCD can be highly localized and robustly controlled, it is considered to be an effective and successful method of controlling TMs. Experiments of ECCD on many devices such as ASDEX Upgrade [10], DIII-D [11], and JT-60U [12] has shown complete suppression of TMs or neoclassical tearing modes, a special class of the tearing modes. Furthermore, numerical studies of TMs stabilization with ECCD have been carried out using various numerical algorithms

[13-17]. However, most of these algorithms are based on the reduced resistive MHD model [18] in slab or cylindrical geometries. Numerical calculations based on the global kinetic/MHD hybrid simulations for more realistic tokamak models are still not available.

The work presented here investigates the stabilization of TMs in plasmas with ECCD. The TM stabilization condition can be achieved in the HL-2A tokamak [19], so this device is considered here as a typical bed for the tearing mode suppression. The HL-2A tokamak (with a major radius of R＝1.64 m and minor radius of a＝0.4 m) is a medium-sized tokamak device. Presently, only TM stabilization by electron cyclotron resonant heating (ECRH) has been achieved. Experimental results show that the global current profile redistribution caused by long pulse ECRH leads to complete mode suppression [20]. Since the TMs and neoclassical tearing mode stabilization by ECCD are still in development, it is essential to conduct the corresponding simulations in order to facilitate and test the design of the real-time control system for TMs. In addition, these computational tools are applied to tearing mode stabilization in DIII-D. R. J. La Haye et al. reported the first use of active feedback to control the the neoclassical tearing mode in DIII-D. [21] Subsequently, the first complete suppression of the m/n = 2/1 tearing mode was achieved using ECCD to replace the 'missing' bootstrap current in the island's O-point [11- 22]. Moreover, the experiments in DIII-D also found that the value of beta is limited by the relation m/n = 2/1 TM in hybrid scenario plasmas [23]. The stationary operation of hybrid plasmas was successfully attained until the ECCD was turned off, suggesting tearing mode stabilization with ECCD is critical for the stable operation.

Presently, there is no systematic theoretical study of TM stabilization using ECCD in HL-2A/M, except for an analytical calculation based on empirical formalism [24]. A. M. Popov et al. have simulated the TM suppression by a radially localized toroidal current from ECCD in DIII-D [25] with full MHD code. In addition, Thomas G. Jenkins has calculated the TM stabilization with ECCD using the NIMROD code and demonstrated the complete suppression of the (2,1) tearing mode [26]. However, their work fell short of realizing a fully coupled, self-consistent model for ECCD/MHD interaction, and the kinetic effect of ion was not accounted for. The kinetic effect of ions on the TM stabilization efficiency present an important area of study that has not been fully investigated, thus motivating our present work. We performed kinetic simulations of tearing modes and their suppression with localized current drive in tokamak plasmas by using the gyrokinetic toroidal code (GTC), [27-28] which has been extensively applied to study neoclassical transport, [29] energetic particle transport, [30] Alfvén eigenmodes, [31-33] microturbulence,[34-35] resonant magnetic perturbations, [36] kink modes, [37] tearing modes [38-39].

The remainder of this paper is organized as follows. The physics model of TM suppression by ECCD is introduced in Sec. 2. The driven current characteristics and its mechanism for controlling the tearing mode as well as the ion kinetic effects on the TM stabilization in HL-2A are presented in Sec. 3. The kinetic simulation of TMs suppression by a radially localized toroidal current from ECCD in DIII-D are described in Sec. 4. Finally, brief conclusions are drawn in Sec. 5.

## 2. Physics model

In order to study the low-frequency MHD instabilities, such as resistive tearing mode, a massless electron fluid model can be coupled with gyrokinetic ions through the gyrokinetic Poisson's equation and Ampere's law [38]. In this work, we neglect the electron kinetic effects, which has be implemented in GTC using a conservative scheme for solving electron drift kinetic equation [40]. To study the effects of ECCD on TM, the ECCD current is obtained by employing ray-tracing and Fokker-Planck equations.

### 2.1 Gyrokinetic ion and massless electron fluid model

To derive the massless electron fluid model, we start with the electron drift kinetic equation. The time evolution of electron guiding center distribution function $f_e$ is given by:

$$\frac{d}{dt} f_e(\mathbf{X}, \mu, v_\parallel, t) = \left[ \frac{\partial}{\partial t} + \dot{\mathbf{X}} \cdot \nabla + \dot{v}_\parallel \frac{\partial}{\partial v_\parallel} \right] f_e = (\frac{\partial}{\partial t} f_e)_{collsion}, \tag{1}$$

where

$$\dot{\mathbf{X}} = v_\parallel \frac{\mathbf{B}}{B_0} + \frac{c\mathbf{b}_0 \times \nabla \phi}{B_0} + \frac{v_\parallel^2}{\Omega_e} \nabla \times \mathbf{b}_0 + \frac{\mu}{m_e \Omega_e} \mathbf{b}_0 \times \nabla B_0, \tag{2}$$

$$\dot{v}_\parallel = -\frac{1}{m_e} \frac{\mathbf{B}^*}{B_0} \cdot \left( \mu \nabla B_0 + q_e \nabla \phi \right) - \frac{q}{m_e c} \frac{\partial A_\parallel}{\partial t}, \tag{3}$$

Here, $\mathbf{X}, \mu, v_\parallel$ is the electron guiding center position, the magnetic moment and parallel velocity. A Krook collisional operator, $(\frac{\partial}{\partial t} f_e)_{collsion} = \eta(f_e - f_{e0})$ is used to provide resistivity, where $f_{e0}$ is the equilibrium distribution function, $m_e$ and $\Omega_e$ are the electron mass and cyclotron frequency, and the expression of $\mathbf{B}^*$ can be found in Ref. [28].

Assuming a shifted Maxwellian for the equilibrium electron distribution function ($f_{e0}$) that satisfies the equilibrium electron drift kinetic equation, Eq.(1) reduces to

$$L_0 f_{e0} = 0, \tag{4}$$

Where $L_0 = \frac{\partial}{\partial t} + (v_\parallel \mathbf{b}_0 + \mathbf{v}_d) \cdot \nabla - \frac{\mu}{m_e} \mathbf{b}_0 \cdot \nabla B_0 \frac{\partial}{\partial v_\parallel}$ is the equilibrium propagator.

Subtracting Eq. (1) by Eq. (4), the equation for the perturbed distribution $\delta f_e$ is

$$L\delta f_e = -\delta L f_{e0} \qquad (5)$$

Where $\delta L = (v_\| \frac{\delta \mathbf{B}}{B_0} + \mathbf{v}_E) \cdot \nabla - (\frac{\mu}{m_e} \frac{\delta \mathbf{B}}{B_0} \cdot \nabla B_0 + \frac{q_e}{m_e} E_\| - \frac{q_e}{m_e} \frac{\mathbf{v}_c}{v_\|} \cdot \nabla \phi) \frac{\partial}{\partial v_\|}$, $L = L_0 + \delta L$.

Defining the particle weight as $w_e = \delta f_e / f_e$, we can rewrite Eq. (1) as the weight equation by using Eq. (5)

$$\begin{aligned}\frac{dw_e}{dt} = (1-w_e)[&-(v_\| \frac{\delta \mathbf{B}}{B_0} + \mathbf{v}_E) \cdot \frac{\nabla f_{e0}}{f_{e0}} \\ &+ (\mu \frac{\delta \mathbf{B}}{B_0} \cdot \nabla B_0 + q_e \frac{B_0^*}{B_0} \cdot \nabla \phi + \frac{q_e}{c} \frac{\partial A_\|}{\partial t}) \\ &\times \frac{1}{m_e} \frac{1}{f_{e0}} \frac{\partial f_{e0}}{\partial v_\|}]\end{aligned} \qquad (6)$$

Integrating Eq. (6), we get the perturbed fluid continuity equation of electron:

$$\begin{aligned}&\frac{\partial}{\partial t}\delta n_e + \mathbf{B}_0 \cdot \nabla(\frac{n_{e0}\delta u_{\|e}}{B_0}) + B_0 \delta \mathbf{v}_E \cdot \nabla(\frac{n_{e0}}{B_0}) \\ &- n_{e0}(\delta \mathbf{v}_{*e} + \delta \mathbf{v}_E) \cdot \frac{\nabla B_0}{B_0} + \delta \mathbf{B} \cdot \nabla(\frac{n_{e0}\delta u_{\|e0}}{B_0}) \\ &+ \frac{c \nabla \times \mathbf{B}_0}{B_0^2} \cdot (-\frac{\nabla \delta P_e}{e} + n_{e0}\nabla \delta \phi) \\ &+ \left\{ \delta \mathbf{B} \cdot \nabla(\frac{n_{e0}\delta u_{\|e}}{B_0}) + B_0 \delta \mathbf{v}_E \cdot \nabla(\frac{\delta n_e}{B_0}) \right. \\ &\left. + \frac{c \nabla \times \mathbf{B}_0}{B_0^2} \cdot \delta n_e \nabla \delta \phi \right\}_{NL} = 0.\end{aligned} \qquad (7)$$

The parallel momentum equation is then equal to:

$$\begin{aligned}&n_{e0}m_e \frac{\partial}{\partial t}\delta u_{\|e} + n_{e0}m_e \mathbf{u}_{\|e0} \cdot \nabla \delta u_{\|e} \mathbf{b}_0 \\ &= -n_{e0}e(-\nabla \delta \phi - \frac{1}{c}\frac{\partial \delta A_\|}{\partial t}) - \frac{\delta \mathbf{B}}{B_0} \cdot \nabla p_{e0} - \nabla_\| \delta p_e - n_{e0}m_e \nu_{ei} \delta u_{\|e}.\end{aligned} \qquad (8)$$

Here, $u_{\|e} = u_{\|e0} + \delta u_{\|e}$, $\delta \phi, \delta A_\|$ are the perturbed electrostatic potential and parallel vector potentials, respectively. Since this work is focused on the resistive tearing mode, we neglect the electron inertial term. Furthermore, we treat the effect of the *rf* source enters as an additional force on the electron fluid, thereby reducing the massless electron momentum equation to the parallel force balance equation:

$$\frac{\partial \delta A_\parallel}{\partial t} = -c\mathbf{b}_0 \cdot \nabla \delta\phi + \frac{c}{n_{e0}e}\mathbf{b}_0 \cdot \nabla \delta p_e - \eta(j - j_{eccd}). \tag{9}$$

Here,

$$j = -\frac{c}{4\pi}\nabla_\perp^2 \delta A_\parallel, \tag{10}$$

with $j_{eccd}$ as the EC-driven current density (see next section), $\eta = m_e \nu_{ei}/e$ as the resistivity, $\nabla_\perp^2 = 1/r(\partial(r\partial/\partial r) - m^2/r^2), \nabla_\parallel = (n - m/q)B_T/RB_0, \quad \delta\mathbf{B} = \nabla \times \delta A_\parallel \mathbf{b}_0$, and $m$ and $n$ as the poloidal and toroidal mode numbers, respectively. Eq. (9) assumes that the ion contribution to the plasma current is small and that the effect of ECCD current on plasma resistivity can be neglected. Dropping the nonlinear term, and considering a uniform equilibrium pressure and the equilibrium current driven term, the Eq. (7) can also be simplified since the drift of electrons (E×B) cancels with that of ions:

$$\frac{\partial}{\partial t}\delta n_e = -\mathbf{B}_0 \cdot \nabla\left(\frac{n_{e0}\delta u_{\parallel e}}{B_0}\right) - \delta\mathbf{B} \cdot \nabla\left(\frac{n_{e0} u_{\parallel e0}}{B_0}\right). \tag{11}$$

In this case we assume the electrons are isothermal along perturbed magnetic field lines, i.e., $T_e$ =constant, $p_e = n_e T_e, \delta p_e = \delta n_e T_e + \delta \mathbf{r} \cdot \nabla(n_{e0} T_e)$. In order to complete the fluid model, using the gyrokinetic Poisson's equation

$$\frac{4\pi Z_i^2}{T_i}(\delta\phi - \delta\tilde{\phi}) = 4\pi(Z_i \delta n_i - e\delta n_e). \tag{12}$$

and the parallel Ampere's law

$$en_{e0}\delta u_{\parallel e} = \frac{c}{4\pi}\nabla_\perp^2 \delta A_\parallel + Z_i n_{i0}\delta u_{\parallel i}. \tag{13}$$

$n_i$ and $u_{\parallel i}$ can be calculated from the standard gyrokinetic model for ions [29-30]:

$$\frac{d}{dt}f_i(\mathbf{X},\mu,v_\parallel,t) = [\frac{\partial}{\partial t} + \dot{\mathbf{X}} \cdot \nabla + \dot{v}_\parallel \frac{\partial}{\partial v_\parallel}]f_i = (\frac{\partial}{\partial t}f_i)_{collision}, \tag{14}$$

$$\dot{\mathbf{X}} = v_\parallel \frac{\mathbf{B}}{B_0} + \frac{c\mathbf{b}_0 \times \nabla\phi}{B_0} + \frac{v_\parallel^2}{\Omega_i}\nabla \times \mathbf{b}_0 + \frac{\mu}{m_i\Omega_i}\mathbf{b}_0 \times \nabla B_0, \tag{15}$$

$$\dot{v}_\parallel = -\frac{1}{m_i}\frac{\mathbf{B}^*}{B_0}\cdot(\mu\nabla B_0 + q_i\nabla\phi) - \frac{q}{m_i c}\frac{\partial A_\parallel}{\partial t}, \tag{16}$$

With $m_i$ and $\Omega_i$ as the ion mass and cyclotron frequency, and the magnetic field $\mathbf{B}^*$ for ions in the same form as in Eqs. (3), where the electron cyclotron frequency is replaced by the ion cyclotron frequency. The collision operator $(\frac{\partial}{\partial t} f_i)_{collision}$ has been implemented in GTC, however, as in Ref. [28], we will omit it in this work.

The fluid electrons (9-11) and the gyrokinetic ions (14-16) are coupled through equations (12) and (13). These equations form a closed system which can simulate the low frequency MHD instabilities.

**2.2 Model for electron cyclotron current drive**

The ECCD current is calculated by the GENRAY/CQL3D software package [41-42]. The two principal equations solved in the package are ray-tracing equations and Fokker-Planck equation.

The ray-tracing equations are:

$$\frac{dR}{dt} = -\frac{c}{\omega}\frac{\partial D_0/\partial N_R}{\partial D_0/\partial \omega}; \frac{dN_R}{dt} = \frac{c}{\omega}\frac{\partial D_0/\partial R}{\partial D_0/\partial \omega}$$
$$\frac{d\varphi}{dt} = -\frac{c}{\omega}\frac{\partial D_0/\partial M}{\partial D_0/\partial \omega}; \frac{dM}{dt} = \frac{c}{\omega}\frac{\partial D_0/\partial \varphi}{\partial D_0/\partial \omega} \quad (17)$$
$$\frac{dZ}{dt} = -\frac{c}{\omega}\frac{\partial D_0/\partial N_Z}{\partial D_0/\partial \omega}; \frac{dN_Z}{dt} = \frac{c}{\omega}\frac{\partial D_0/\partial Z}{\partial D_0/\partial \omega}$$

Here we use cylindrical coordinates $\boldsymbol{R}=(R, \varphi, Z)$, where $R$ is the major radius, $\varphi$ is the toroidal angle, and $Z$ is along the vertical axis. $\boldsymbol{N}=\boldsymbol{K}c/\omega=(N_R, M=RN_\varphi, N_Z)$. In the code, the poloidal injection angle, $\alpha$, is defined with respect to the $Z$ = constant plane at the source, with positive angles above the plane and negative below. The toroidal injection angle, $\beta$, is measured counterclockwise with respect to the $Z$ axis. $\omega$ denotes wave frequency and $D_0$ the dispersion relation calculated with cold plasma approximation.

The 3-dimensional bounce averaged Fokker-Planck equation, 2-D in momentum space (slowing down, pitch-angle) and 1-D in configuration space (radial dimension) for the electron distribution function $f_e$ are given by:

$$\frac{\partial f_e}{\partial t} = \frac{\partial f_e}{\partial \boldsymbol{u}} D_{EC} \frac{\partial f_e}{\partial \boldsymbol{u}} + \hat{C} f_e, \quad (18)$$

where $D_{EC}$ is the diffusion coefficients of the electron cyclotron wave (ECW) in the velocity space, $\boldsymbol{u} = \boldsymbol{p}/m_e = \gamma \boldsymbol{v}/c$ is the normalized momentum, and $\hat{C}$ is the collision operator. Solving the Fokker-Planck equation, we obtain the distribution function. The driven current density can be calculated from:

$$j_{eccd}(r) = -en_{e0}c\int \frac{u_\parallel}{\gamma} f_e(r, \boldsymbol{u})d\boldsymbol{u}. \quad (19)$$

Once the ECCD current is obtained, the effect of the ECCD on tearing mode can be implemented through Eq. (9). If we use Eq.(19) as $j_{eccd}$ in Eq. (9) directly, it means that we consider the cases with the continuous current drive. It is well known that the driven current at the O-point can suppress the tearing mode, while at the X-point, it leads to the destabilization of the tearing mode [43]. In experiment, helical current has been generated by modulating continuous current drive, and better efficiency for suppressing tearing modes has been obtained [44]. Therefore, a helical driven current is also used to study the suppression of tearing modes. This kind of driven current density can be written as follows:

$$j_{eccd} = j_{eccd}(r)[1+\cos(m\theta+n\zeta)] \qquad (20)$$

It should be noted that due to computational limitations, Eqs. (17)-(19) in GENRAY/CQL3D package are not solved simultaneously with Eqs. (1-3) and (9-15) in GTC, thus, the effect of the magnetic perturbations on the wave deposition and the source current profile is neglected.

### 3. GTC simulations of tearing mode instability in HL-2A-like equilibrium
### 3.1 Linear simulation of tearing mode without current drive

Firstly, an HL-2A-like equilibrium is chosen in our simulation, i.e., the major radius is $R_0 = 1.65$ m and the minor radius is $a = 0.4$ m. The initial equilibrium $q$ profile (two $q$ profiles are presented: experimental $q$ from the EFIT equilibrium reconstruction [45], the experiment-like $q$ used in simulations), the electron density, and the temperature profile are shown in Figure 1. The $q=2$ rational surface of the two $q$ profiles has different positions and magnetic shear. Our calculation shows the $q$ we use in simulation enhance the tearing mode, whereas the experimental $q$ gives rise to a damped TM. Moreover, as we shift the experimental $q$ profile from the equilibrium reconstruction inward, the TM is still damped. Therefore, it is the magnetic shear that leads to increasing tearing mode in HL-2A. Since the experimental $q$ leads to TM stabilization, the experiment-like $q$ profile (the red line in Fig. 1) will be used in the subsequent calculations. The magnetic field $\mathbf{B}$ and the current density $J$ is calculated with the EFIT code. In the simulations, we use number of grids 150×350×16 in the radial, poloidal and parallel direction respectively. The equilibrium plasma current is 157 kA, the $q=2/1$ surface is $r=0.6a$, and the plasma resistivity is $\eta = 1.0\times10^{-5}\,\Omega/m$.

It is important to note that the resistivity is higher than the Spitzer resistivity (~$10^{-8}\Omega$/m with the HL-2A parameter). This is due to the very large time steps that are required, causing numerical imprecision in the finite mesh fluid model when the resistivity is very low, and preventing us to achieve $\eta < 10^{-8}\Omega$/m in the present version of the GTC. However, we have calculated the dependence of the linear growth rate on resistivity and found that the dependence is similar to the theoretical resistivity scaling of tearing modes, i.e., $\eta^{3/5}$. Figure 2 shows the radial mode structure from eigenvalue calculation and GTC simulation, respectively. It can be seen that the GTC fluid simulation result agrees well with the eigenvalue result. The growth rates from GTC and eigenvalue methods are $0.0040\omega_A$ and $0.0036\omega_A$, respectively, where $\omega_A$ is the Alfvén frequency. Figure 3 is the mode structures of the parallel vector potential $\delta A_\parallel$ and the electrostatic potential $\delta\phi$, and the island sketch map at $t = 2.2 \times 10^{-5} s$ on the poloidal plane. The mode amplitude oscillates in the early stage and then starts to increase linearly at $t = 1.5 \times 10^{-5} s$. The corresponding magnetic island width is about $0.168a$ at this time, and the linear growth rate is $0.13\omega_s$, where the normalized frequency is $\omega_s \equiv c_s/R_0, c_s \equiv \sqrt{T_e/m_i}$.

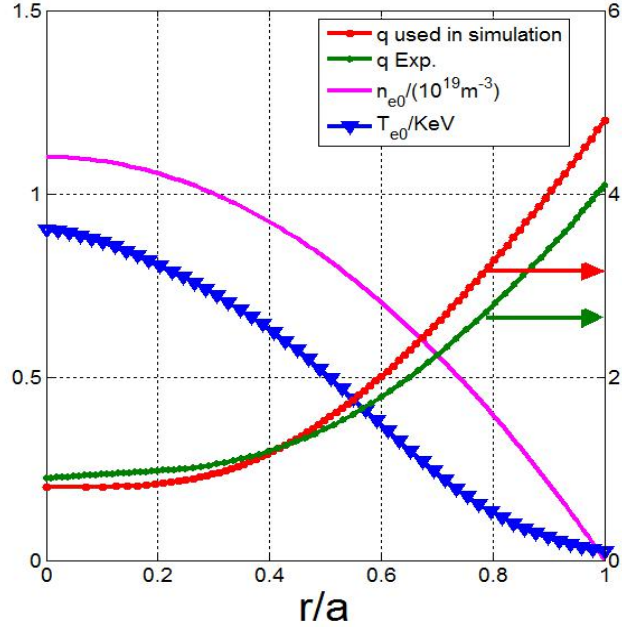

Figure 1. Radial profiles of the safety factor, electron density and temperature.

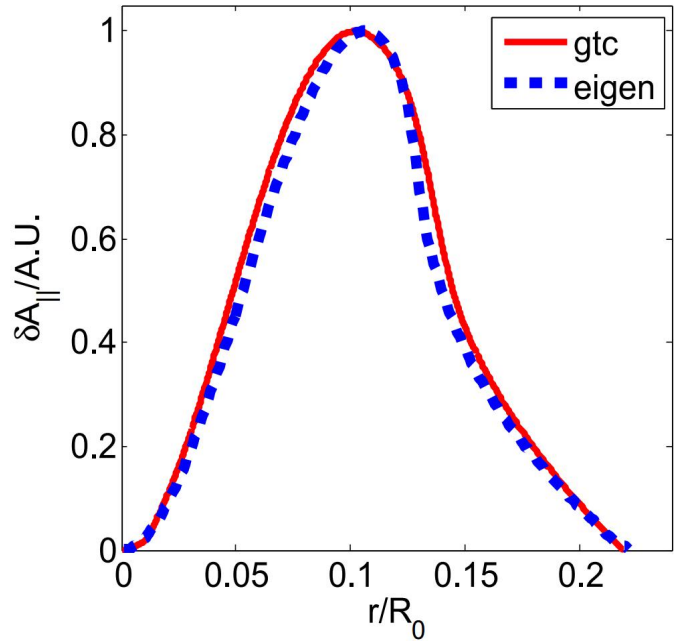

Figure 2. Comparison of the radial mode structures of (2, 1) tearing mode from GTC simulation and 1D eigenvalue calculation in the cylindrical geometry.

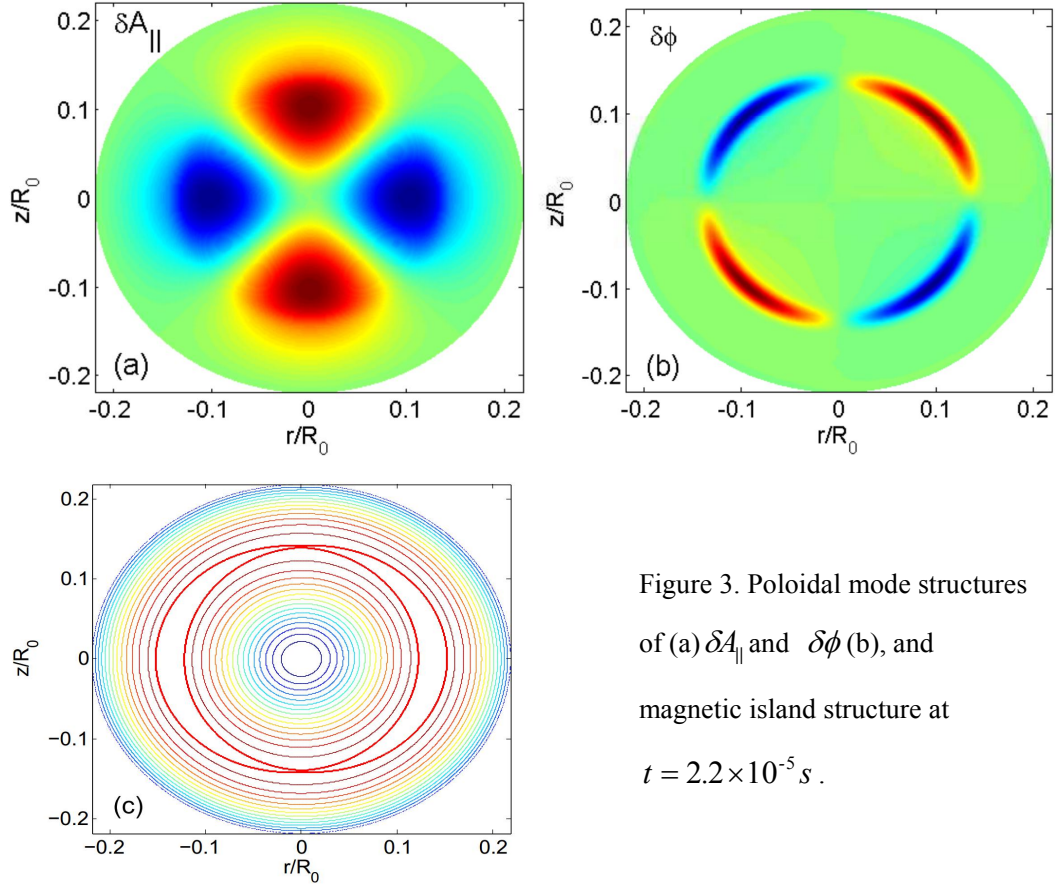

Figure 3. Poloidal mode structures of (a) $\delta A_\parallel$ and $\delta\phi$ (b), and magnetic island structure at $t = 2.2\times 10^{-5}\,s$.

### 3.2 Tearing mode instability with stationary ECCD

A typical electron cyclotron current drive in HL-2A -like equilibrium is shown in Figure 4. Figure 4 (a) shows the EC-wave trajectories, and Fig. 4 (b) shows the current density versus normalized minor radius. In this case, the poloidal injection angle is $\alpha=113°$, the toroidal injection angle is $\beta=190°$, the wave power is 1MW, and the wave frequency is 68 GHz X2-mode. The total driven current here is 13 kA (the current ratio CR≡$I_{rf}$ /$I_0$=8%), with the radial deposition position located at r/a=0.6, which is approximately at the rational surface of $q$=2.

The poloidal profiles of the equilibrium current density and continuous ECCD current density corresponding to Fig. 3 are shown in Fig. 5. Presented in Fig.6 is the helicon ECCD current density in poloidal cross section. An example of the evolutions of tearing mode magnetic island width without ECCD, with continuous ECCD, and helicon ECCD is shown in Fig. 7. It can be seen that without ECCD, (2,1) TM increases linearly after an initial oscillating phase. In the case of ECW injection, width of the TM island decreases quickly and reduces to zero at about $t=3.5\times10^{-5}$s. Moreover, the growth rate of TM is negative (about $-0.10\omega_s$), thus, the TM is indeed damped by ECCD. Finally, the helicon current drive is more efficient than the continuous ECCD. In comparison with continuous ECCD, the growth rate with helicon current drive is lower, and about $-0.25\omega_s$. In summary, we find that 1MW ECW is sufficient to suppress the (2,1) tearing mode in a typical HL-2A equilibrium.

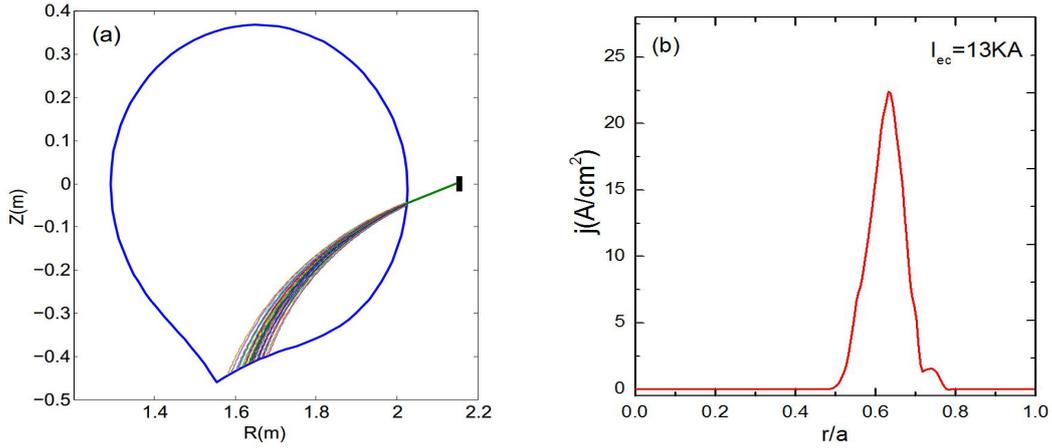

Figure 4. (a) EC-wave trajectories, (b) current density versus normalized minor radius.

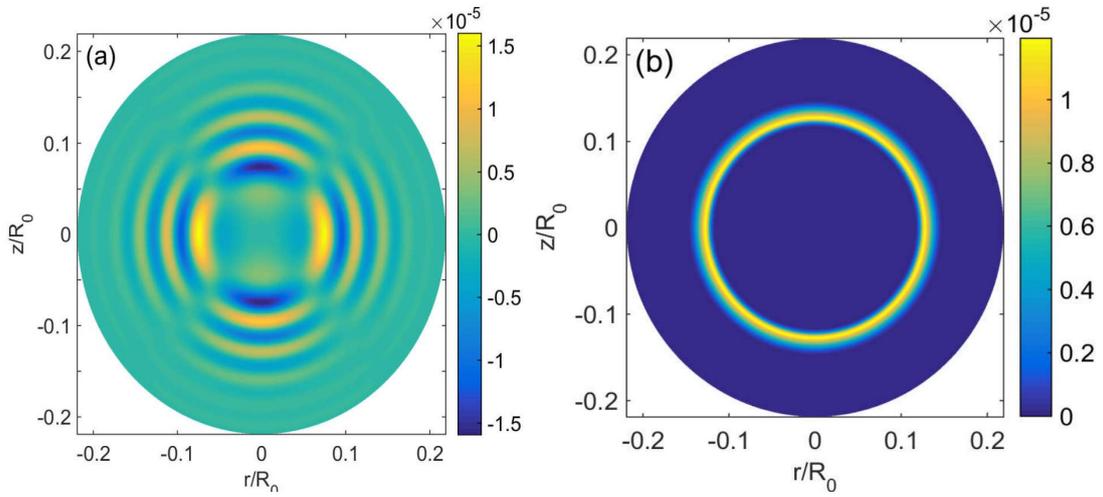

Figure 5. The poloidal structure of perturbed current (a) and ECCD current (b).

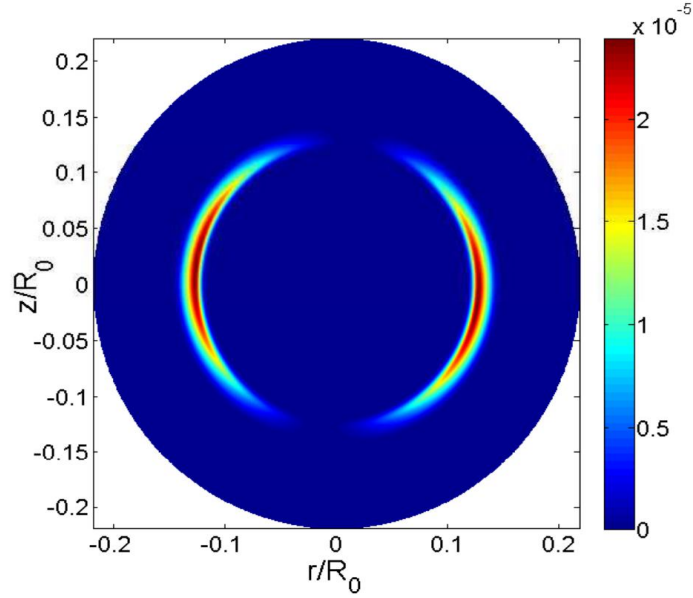

Figure 6. The poloidal structure of helicon ECCD current.

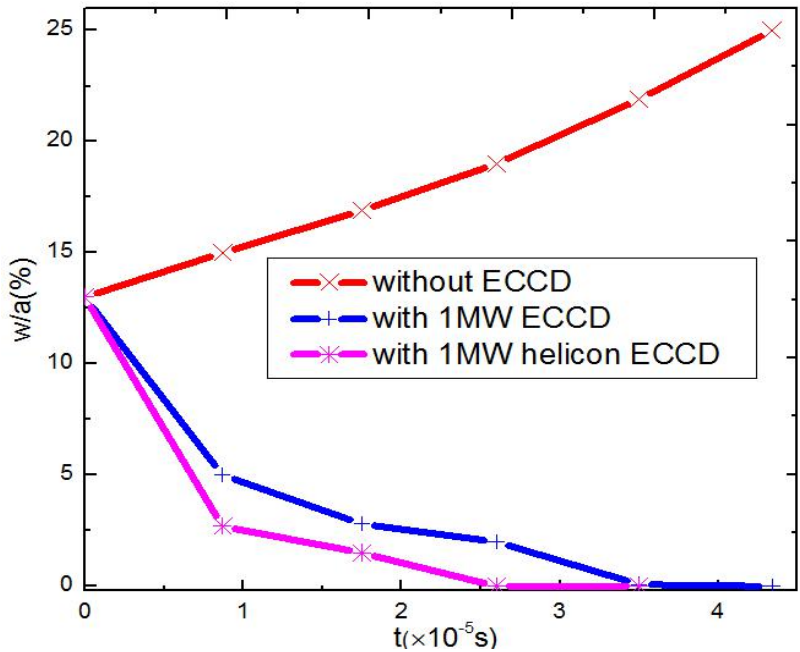

Figure 7. Time evolution of the width of tearing mode magnetic island without, with continous ECCD, and helicon ECCD.

In general, the tearing mode in HL-2A is straightforward to suppress for this equilibrium parameter. Since the steering mirrors in the launcher allow the poloidal injection angle and the toroidal injection angle to be rotated to -20°~20°, it is possible to inject the ECW off-axis, i.e., r/a=0.6 in this case. The TM stabilization is closely related to the value of the current ratio CR. A previous study [39] found that the tearing mode can be completely suppressed when the current ratio CR is about 4%. The equilibrium plasma current $I_0$ is low in this case. However, the total ECW power of 3MW [46] is still sufficient to suppress the TMs, even if $I_0$ reaches the highest value in HL-2A (450 kA). Finally, the dependence of TMs magnetic island width and growth rates on wave misalignment has been investigated in Ref. [39].

### 3.3 Ion kinetic effects on the tearing mode stabilization

In high-temperature plasmas, the kinetic effects of ions can have a significant effect on the evolution of tearing modes. Analytically, Cai et al. found that the co-circulating energetic ions can stabilize the TMs [9]. Subsequently, they reported the effects of energetic particles on TMs from M3D-K code simulations [47]. However, in their model, background thermal ions and electrons are treated as a single fluid, and the energetic ions are described by the kinetic drift equation. However, in our model, only the electrons are treated as fluid, while both background thermal ions and energetic ions are treated kinetically. Therefore, we can study the kinetic effects of thermal ions effectively using the GTC code.

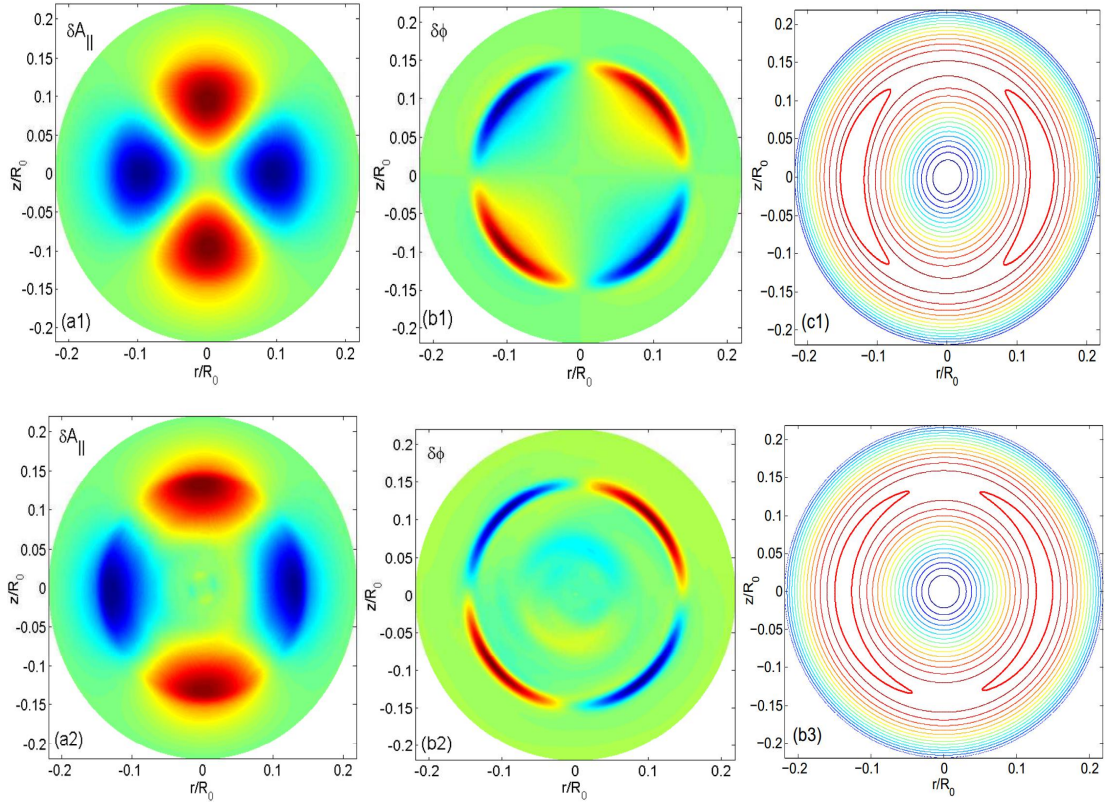

Figure 8. The mode structure (a,b), the magnetic island of tearing mode (c), without thermal ions (top panel), and with thermal ions (the bottom panel).

In our simulation, the number of grids $150 \times 350 \times 16$ in the radial, poloidal and parallel direction are still used. We adopt the same HL-2A parameters as in Section 3.1, and load 200 0000 ions for the kinetic calculation. Figure 8 shows the mode structure (a and b) and the magnetic island of TMs (c) without kinetic thermal ions (top), and with kinetic thermal ions (bottom). As can be seen in Fig. 8, both $\delta A_\parallel$ and $\delta\phi$ shrink in the case without kinetic thermal ions. Figure 8(c) shows a clear difference in island width in these two cases. The island radial width of the TMs with kinetic ions is significantly smaller than that without kinetic ions at the same time, $t = 1.9 \times 10^{-5} s$. The growth rate decreases by about 0.12 $\omega_s$ when the ion kinetic effects are included. The growth rate becomes negative when injecting ECW power, and is about -0.29 $\omega_s$ considering both the thermal ions and ECW power, indicating that the kinetic effects of thermal ions enhance the TM stabilization for electron cyclotron wave injection.

Moreover, with the ions added, we can see that there is a weak rotation of modes structure in clockwise, which is in the same direction of ion diamagnetic drift direction. This rotation is more prominent when we calculate a case with $q=1.75+4.66r^2$, where the $q=2$ rational surface lies to more inner plasma. And the calculated rotation frequency is 1.08KHz, which is very small in comparison with the ion diamagnetic drift frequency, $\omega_i^*$, $\omega_i^* = \dfrac{T_i}{e} \dfrac{dln(p_i)}{d\psi_p} n = 7.9 kHz$.

Figure 8 has shown that the radial mode width shrinks when kinetic thermal ions are added, and the net effect of kinetic particles on tearing modes is significant and stabilizing. We further scan the temperature of thermal ions in order to study the kinetic effects of thermal ions. Figure 9 shows the dependence of ion temperature for the island width on the poloidal plane and the growth rates of TM. It can be seen that the island radial width increases with the decrease of ion temperature. The growth rate for the three cases, $T_i$= 0.01, 0.1, and $1T_e$, are $0.10\omega_s$, $-0.09\omega_s$, $-0.12\omega_s$, respectively. It can be concluded that with the decrease of the ion temperature, both the island radial width and the growth rates approach the values from fluid simulations.

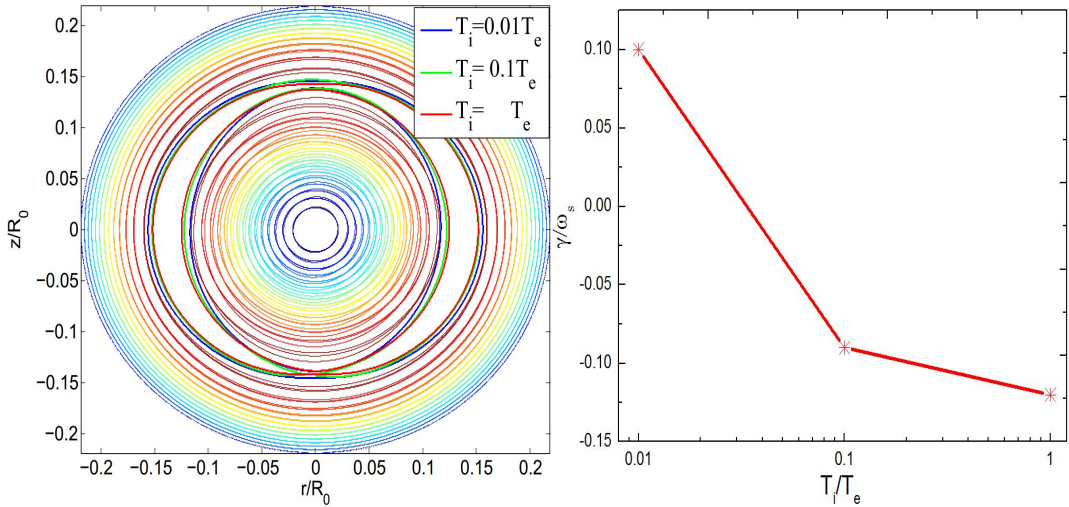

Figure 9. Island map on the poloidal plane with different thermal ion temperature with total beta value the same for the three cases(the left).Growth rates of TM versus the ion temperature (the right) .

## 4. GTC simulations of tearing mode instability for DIII-D

As mentioned before, the performance of plasmas in DIII-D can be limited by m/n = 2/1 tearing modes [23,48], and continued island growth due to neoclassical tearing modes that can lead to disruptions. However, stability could be improved by the ECCD. Stationary stable operation at high beta was successfully sustained until the ECCD was turned off and the expected m/n = 2/1 mode appeared. The mechanism of the TMs stabilization by ECCD should be theoretically elucidated. This simulation of the tearing mode stabilization in DIII-D is presented in the following part.

### 4.1 Linear simulation of tearing mode without current drive

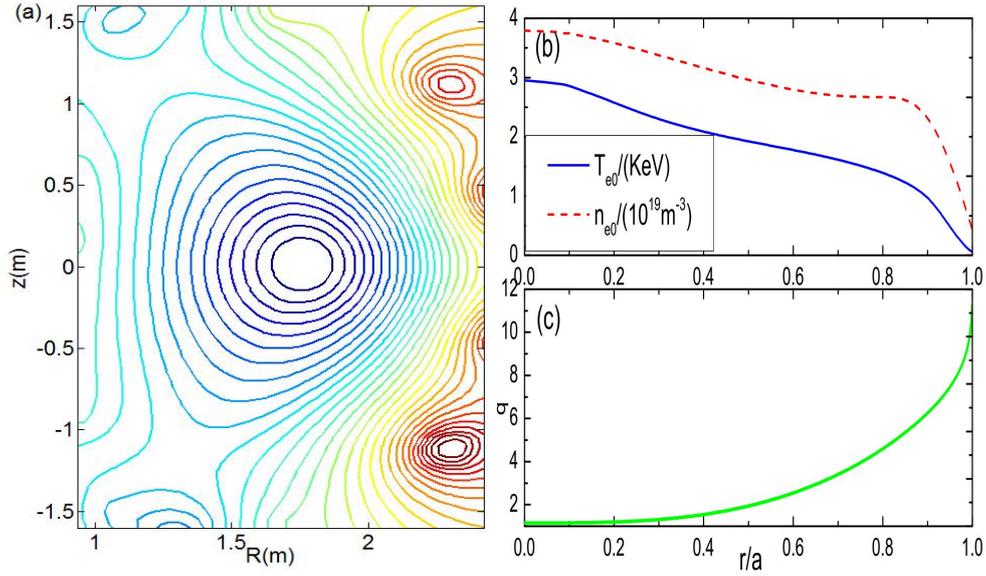

Figure 10. (a) Poloidal magnetic structure of DIII-D equilibrium, (b) radial profiles of electron density and temperature, (c) safety factor.

Firstly, the equilibrium of DIII-D discharge 157402 is used. This discharge has a prominent neoclassical tearing mode. However, this work is concentrated on the TM simulation and its suppression, therefore, we consider only the 2D equilibrium profiles. The discharge parameters are given as follows: the major radius is $R_0 = 1.78$ m, the minor radius is $a=0.58$m, the equilibrium plasma current is 790 kA, the q=2/1 surface is $r=0.6a$, and the toroidal magnetic field strength is $B_T = 2.06$ T. In the simulations here, we use number of grids 128×512×32 in the radial, poloidal and parallel direction respectively. The configuration, initial equilibrium q profile, the electron density, and the temperature profile are shown in Figure 10. The resistivity is set to $\eta = 1.0\times10^{-5}\,\Omega/m$. Figure 11 shows the mode structures of the parallel vector potential $\delta A_\parallel$ and electrostatic potential $\delta\phi$, and the island structure at $t = 2.7\times10^{-5}\,s$ on the poloidal plane. The mode amplitude increases linearly at about $t = 5.5\times10^{-5}\,s$, with the corresponding magnetic island width of about $0.13a$ at this time and the linear growth rate of $0.016\omega_s$. Therefore, Fig. 11 shows the (2, 1) tearing mode structures in DIII-D configuration. It should be noted that the magnetic island width given here is dependent on the initial field perturbation, which is given by $-2.632 \times 10^{-3}(r/R_0)^2(1-r/R_0)^2$. Our simulations show that tearing mode is unstable in this DIII-D discharge, which may provide the seed island for the neoclassical tearing mode.

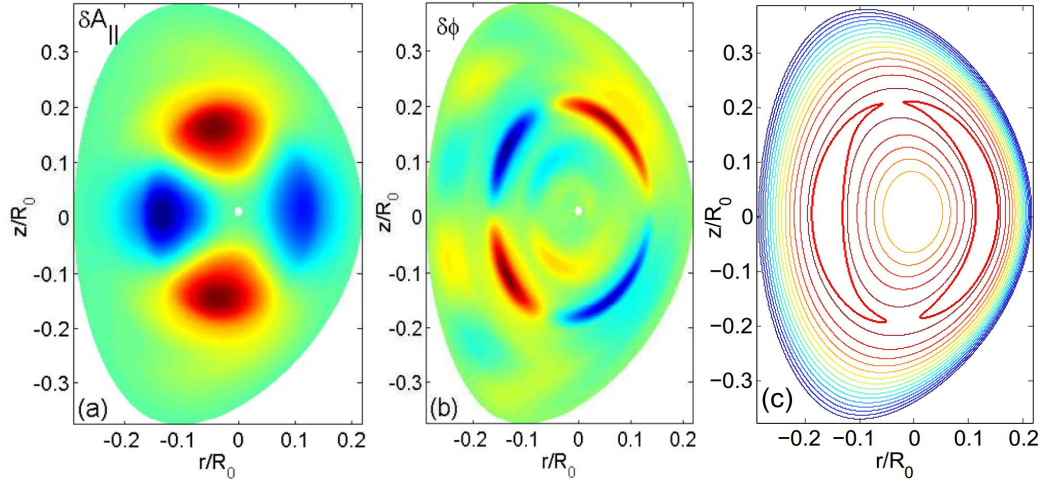

Figure 11. Poloidal mode structures $\delta A_\parallel$ (a) and $\delta\phi$ (b), and the island structure at $t=2.8\times10^{-5}$s (c).

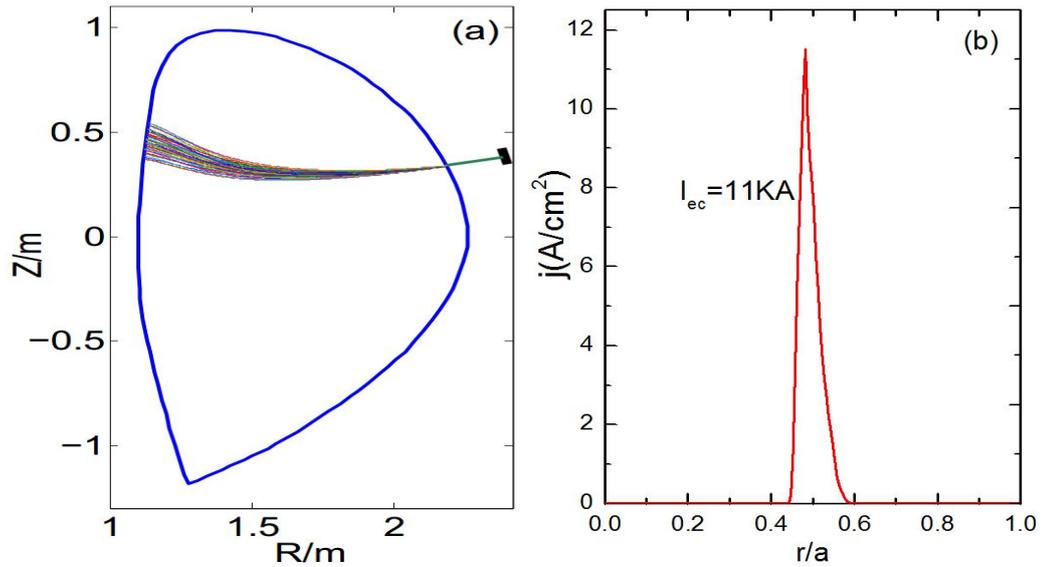

Figure 12. (a) EC-wave trajectories, (b) current density versus normalized minor radius in DIII-D tokamak.

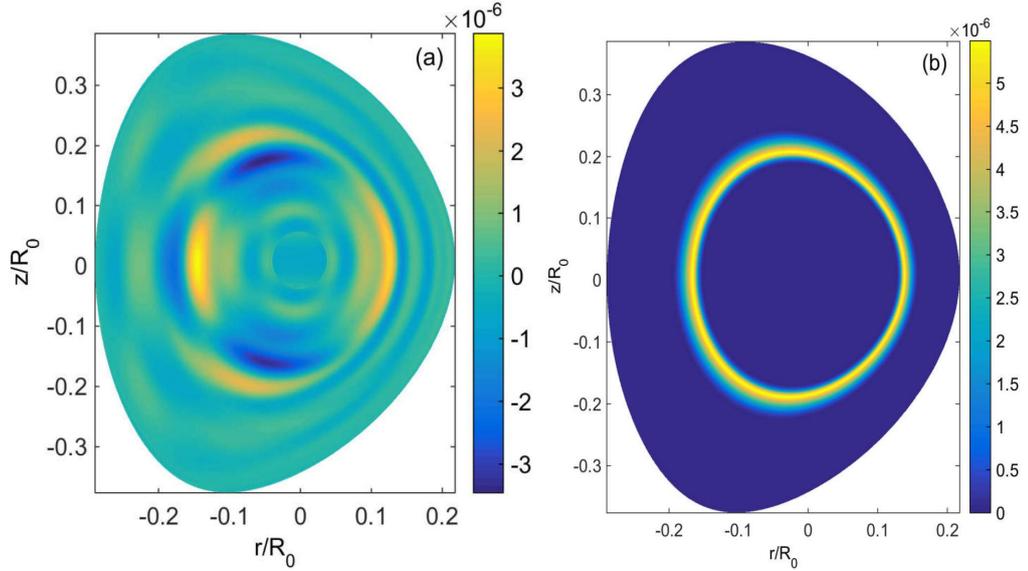

Figure 13. (a) Poloidal structure of perturbed current, (b) EC-driven current DIII-D tokamak.

**4.2 Tearing mode instability with stationary ECCD**

For equilibrium and plasma parameters used in Section 3.1, a 110 GHz electron cyclotron wave is launched with X-mode polarization from a port above the midplane, with the trajectories of the electron cyclotron wave shown in Figure 12 (a). With 1MW ECW power injected, the corresponding profile of driven current density with a poloidal angle of 100° and a toroidal angle of 193° is shown in Figure 12 (b). The total driven current is 11 kA (CR=1.3%), with the radial deposition position located at r/a=0.5 and a very narrow current drive profiles characteristic of ECCD, about 3.5 cm full width half maximum (FWHM), which is well suited for stabilizing TMs.

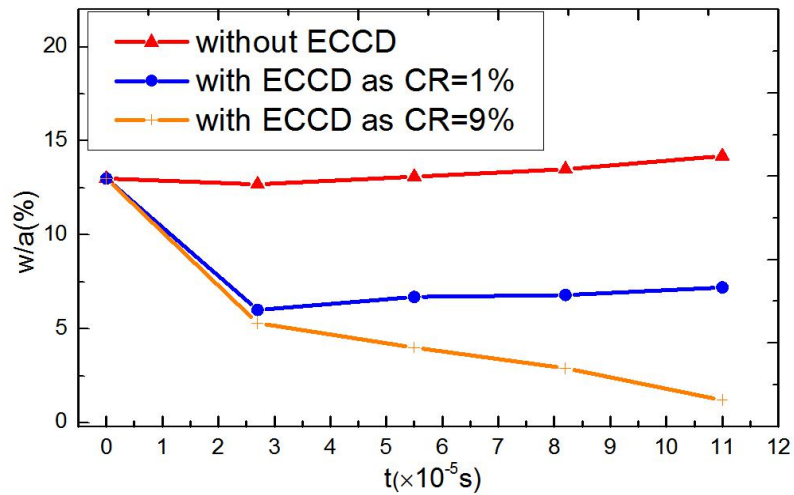

Figure 14. Time evolution of the width of tearing mode magnetic island with and without the ECCD.

Figure 13 shows the poloidal profiles of the equilibrium current density and ECCD current density corresponding to Fig. 10. An example of the evolution of a TM magnetic island width with and without the ECCD in Fig. 12(b) is shown in Figure 14.

Without ECCD, the TMs grow slowly and form a linear eigenmode at $t = 5.5 \times 10^{-5} s$.

The stability is improved with ECCD since the magnetic island width decreases with ECCD, however, the island width does not go to zero in this case. The ECCD only reduces the growth the TMs, rather than fully stabilizing it. We surmise that this is caused by the low ECCD current (11 kA), significantly below the equilibrium current (790 kA), and that a higher input ECW power is required to entirely suppress the TMs. Therefore, if the value of CR is increased, the tearing modes are almost suppressed when CR equals to 9% (as depicted by the orange line in Fig. 14). The corresponding growth rate is -0.08$\omega_s$.

**4.3 Ion kinetic effects on the tearing mode stabilization**

Similar to the kinetic simulation in HL-2A, we initiate the GTC kinetic simulations by loading 200 0000 ions in the DIII-D configuration. Figure 15 shows the mode structure (a and b) and the magnetic island of tearing mode (c) without kinetic thermal ions (top), and with kinetic thermal ions (bottom). This figure also shows the difference of $\delta A_\parallel$ and the $\delta \phi$ with and without kinetic thermal ions. The radial width of the TM island with kinetic thermal ions is also significantly smaller than that without kinetic thermal ions at $t = 0.68 \times 10^{-5} s$.

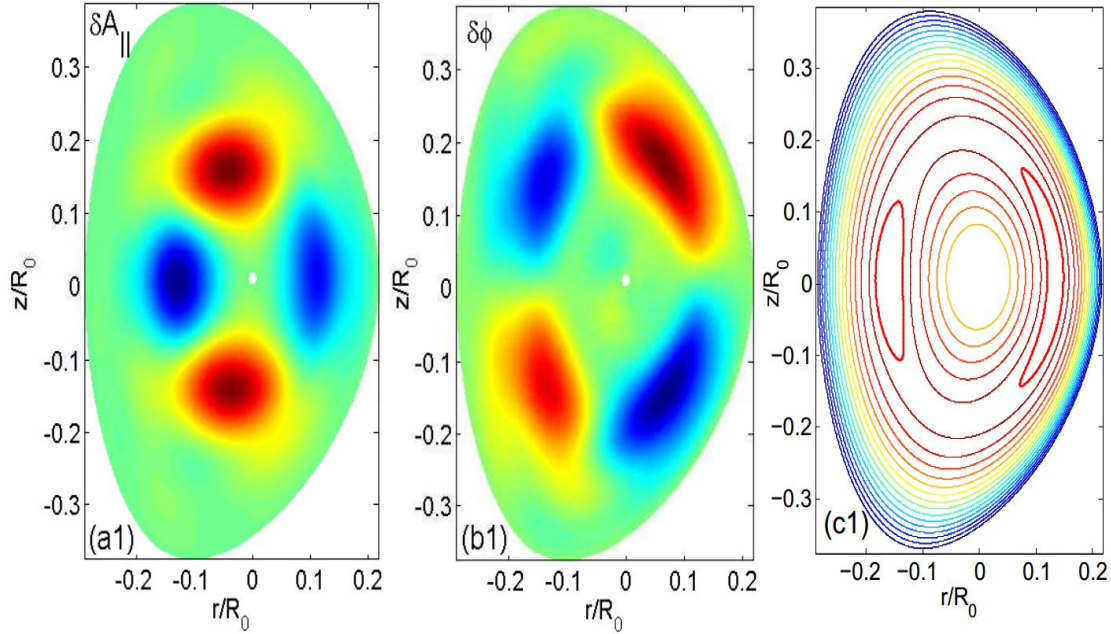

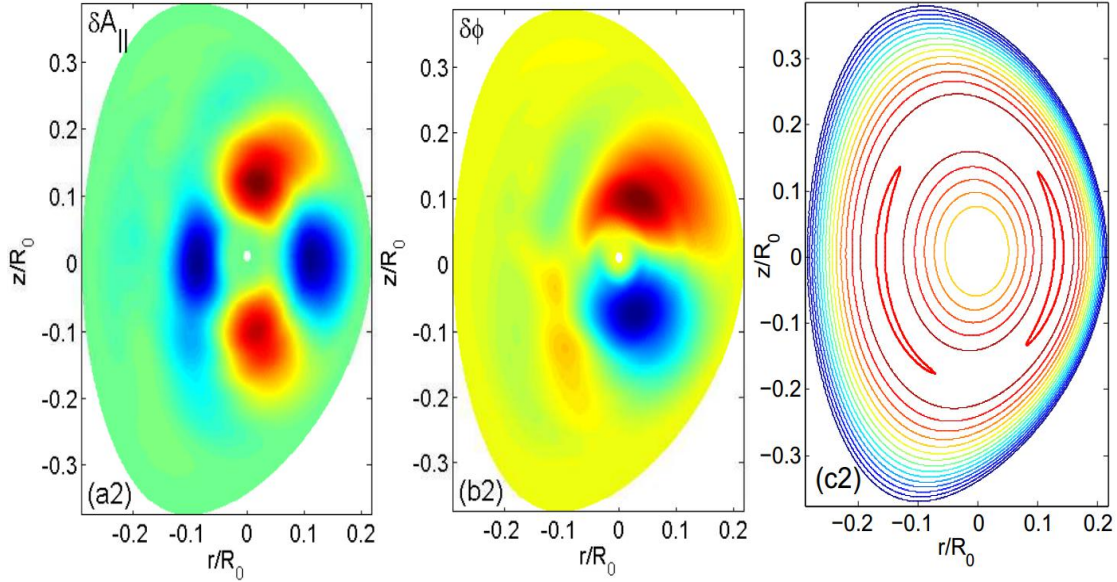

Figure 15. Poloidal mode structure and the magnetic island of tearing mode without (top panel) and with kinetic thermal ions (bottom panel) in DIII-D.

As can be seen, both in HL-2A and DIII-D, kinetic effects of thermal ions can reduce the island width, hence reducing the growth rate of the TMs in a tokamak plasma. This result is consistent with analytical calculations, indicating that the kinetic ion effects tend to reduce the degree of singularity of the solution in the outer region where the plasma is normally well described by ideal MHD equations. It should be noted that both the simulation in HL-2A and in DIII-D configuration is performed in toroidal geometry, and that the kinetic effects of thermal ions are observed. Therefore, the stabilizing effect of kinetic ion is dominant over the destabilizing effect in all our simulations.

## 5. Conclusions

In summary, we have investigated the influence of the electron cyclotron current drive on the m/n=2/1 tearing mode using gyrokinetic simulations in a HL-2A-like equilibrium and DIII-D configuration. The tearing mode evolution is calculated with a gyrokinetic ion and massless electron fluid model, and the rf current source is obtained obtained by ray-tracing and the Fokker-Planck method. The TMs are found to be perfectly stabilized by a continuous 1MW 68GHz X2-mode in HL-2A tokamak. And the helicon current drive is more efficient than the continuous ECCD. While the (2/1) tearing mode in the DIII-D tokamak is only partially stabilized with the 1MW 110GHz X2-mode due to the inadequate power input. Generally speaking, the tearing mode in HL-2A is easy to be suppressed with the present control system.

The ion kinetic effect on the tearing mode stabilization is demonstrated. Analysis of the GTC simulation reveals, both in HL-2A and DIII-D, that the presence of ions can reduce the island width as well as the growth rate due to the interaction between the ions and the TMs, hence benefit to the TMs stabilization with ECCD. With the ions added, we can see that there is a weak rotation of modes structure in clockwise. At the same time, the kinetic effect of thermal ions on TM is found to be more pronounced with higher ion temperature. Our simulations under a certain machine configuration will contribute to the design of the real-time control system of the TMs, and provide useful suggestions to nearby TMs or neoclassical tearing mode control experiments for fusion device, especially for HL-2A and DIII-D tokamaks. Meanwhile, the calculations will promote to the longer term plan of building a first principles model and a self-consistent simulation of the neoclassical tearing mode in fusion plasmas.

**Acknowledgements**

The authors would like to thank Min Xu, and L. W. Yan, Jun Wang at SWIP, Lei Shi and Jian Bao at UCI, Youjun Hu at ASIPP for fruitful discussions. J. C. Li would like to thank S. Y. Liang at SWIP, X. J. Tang at PKU for providing the experimental data in HL-2A and DIII-D. This work is supported by the ITER-China program (2014GB107004, 2013GB111000), NSFC (11375053), U.S. SciDAC GSEP, and the China Postdoctoral Science Foundation No. 2017M610023.

**References**

[1] Giruzzi G et al 1999 *Nucl. Fusion* **39 107**.
[2] Hegna C C 1998 *Phys. Plasmas* **5 1767**.
[3] La Haye R J 2006 *Phys. Plasmas* **13 055501**.
[4] Lin-Liu Y R et al 2003 *Phys. Plasmas* **10 4064**.
[5] Pinsker R I , 2001 *Phys. Plasmas* **8 1219**.
[6] Fitzpatrick R 1993 *Nucl. Fusion* **33 1049**.
[7] Hu Q et al 2012 *Nucl. Fusion* **52 083011**.
[8] Hegna C C et al 1989 *Phys. Rev. Lett*. **63 2056**.
[9] Cai H et al 2011 *Phys. Rev. Lett.* **106 075002**.
[10] Gantenbein G et al 2000 *Phys. Rev. Lett.* **85 1242**.
[11] Petty C C et al 2004 *Nucl. Fusion* **44 243**.
[12] Isayama A et al 2001 *Nucl. Fusion* **41 761**.
[13] Yu Q et al 2000 *Phys. Plasmas* **7 312**.
[14] Yu Q et al 2004 *Phys. Plasmas* **11 1960**.
[15] Jenkins T G et al 2015 *J. Comput. Phys.* **297 427**.
[16] Chen L et al, 2014 *Phys. Plasmas* **21 102106**.
[17] Wang S et al 2016 *Physics of Plasmas* **23 459**.
[18] Comisso L et al 2010, *Nucl. Fusion* **50 125002**.
[19] Duan X R et al 2017 2017 Nucl. Fusion **57 102013.**
[20] Liu Y et al. 2008 Study on stabilization of tearing mode with ECRH and its resultant transport properties on HL-2A tokamak. *Geneva: IAEA Fusion Energy Conference* **EX-/P9-2**.
[21] Haye R J L et al 2002 *Physics of Plasmas* **9 2051**.
[22] Prater R et al 2007 *Nucl. Fusio* **47 371**.
[23] La Haye R J et al 2008 *Nucl. Fusion* **48 015005.**
[24] Li J C et al 2016 *Chinese Physics B* **25 214**.

[25] Popov A M et al 2002 *Phys. Plasmas* **9 4229**.
[26] Jenkins T G et al 2010 *Phys. Plasmas* **17 012502.**
[27] Lin Z et al 1998 *Science* **281 1835.**
[28] Holod I et al 2009 *Phys. Plasmas* **16 122307**.
[29] W. W. Lee 1987 *J. Comput. Phys*. **72 243**.
[30] A. Brizard et al 2007 *Rev. Mod. Phys*. **79** 421.
[29] Zhang W et al 2008 *Phys. Rev. Lett*. **101 095001**.
[30] Lin Z et al 1997 *Phys. Rev. Lett*. **78 456**.
[31] H S Zhang et al 2012 *Phys. Rev. Lett*. **109 025001**.
[32] Wang Z et al 2013 *Phys. Rev. Lett*. **111 145003.**
[33] Liu Y et al 2017 *Nucl. Fusion* **57 016005.**
[34] Lin Z et al 2007 *Phys. Rev. Lett*. **99 265003.**
[35] Xiao Y et al 2009 *Phys. Rev. Lett*. **103 085004**.
[36] Holod I et al 2017 *Nucl. Fusion* **57 016005.**
[37] McClenaghan J et al 2014 *Phys. Plasmas* **21 122519.**
[38] Liu D J et al 2014 *Phys. Plasmas* **21 122520.**
[39] Li J C et al 2017 *Physics of Plasmas* **24 082508.**
[40] J. Bao, D. Liu, Z. Lin, Phys. Plasmas 24, 102516 (2017).
[41] Harvey R W et al 2001 The GENRAY Ray Tracing Code *CompX Report* **CompX-2000–01.**
[42] Harvey R W et al 1992 (Proceedings of IAEA TCM on Advances in Simulation and Modeling of Thermonuclear Plasmas (Montreal, Canada, 1992).
**[43]** Zhang W et al 2017 *Phys. Plasmas* **24 062510.**
[44] Maraschek M et al 2007 *Phys. Rev. Lett*. **98 025005**.
[45] Lao L et al 1985 *Nucl. Fusion* **25 1611**.
[46] Li J C et al 2015 *Phys. Plasmas* **22 062512**.
[47] Cai H et al 2012 *Phys. Plasmas* **19 072506**.
[48] La Haye R J et al 2006 *Nucl. Fusion* **46 451.**